\documentclass[manuscript]{aastex}

\usepackage{graphicx}
\usepackage{epsfig}
\usepackage{amsmath}
\usepackage{natbib}
\usepackage{subfigure}
\usepackage{here}

\begin{document}

\title{${\it Hinode}$ and ${\it IRIS}$ observations of the magnetohydrodynamic waves propagating from the photosphere to the chromosphere in a sunspot}

\author{Ryuichi Kanoh \altaffilmark{1,2}, Toshifumi Shimizu \altaffilmark{2,1} , and Shinsuke Imada \altaffilmark{3}}

\affil{\altaffilmark{1} Department of Earth and Planetary Science, The University of Tokyo, 7-3-1 Hongo, Bunkyo-ku, Tokyo 113-0033, Japan }
\affil{\altaffilmark{2} Institute of Space and Astronautical Science, Japan Aerospace Exploration Agency, 3-1-1 Chuo-ku, Sagamihara, Kanagawa 252-5210, Japan}
\affil{\altaffilmark{3} Institute for Space-Earth Environmental Research, Nagoya University, Furo-cho, Chikusa-ku, Nagoya, Aichi 464-8601, Japan}
\email{kanoh.ryuichi@ac.jaxa.jp}
\email{shimizu@solar.isas.jaxa.jp}
\email{shinimada@stelab.nagoya-u.ac.jp}

\begin{abstract}
Magnetohydrodynamic (MHD) waves have been considered as energy sources for heating the solar chromosphere and the corona. Although MHD waves have been observed in the solar atmosphere, there are a lack of quantitative estimates on the energy transfer and dissipation in the atmosphere. We performed simultaneous {\it Hinode} and {\it IRIS} observations of a sunspot umbra to derive the upward energy fluxes at two different atmospheric layers (photosphere and lower transition region) and estimate the energy dissipation. The observations revealed some properties of the observed periodic oscillations in physical quantities, such as their phase relations, temporal behaviors, and power spectra, making a conclusion that standing slow-mode waves are dominant at the photosphere with their high-frequency leakage, which is observed as upward waves at the chromosphere and the lower transition region. Our estimates of upward energy fluxes are $2.0\times10^7$ erg cm$^{-2}$ s$^{-1}$ at the photospheric level and $8.3\times10^4$ erg cm$^{-2}$ s$^{-1}$ at the lower transition region level. The difference between the energy fluxes is larger than the energy required to maintain the chromosphere in the sunspot umbrae, suggesting that the observed waves can make a crucial contribution to the heating of the chromosphere in the sunspot umbrae. In contrast, the upward energy flux derived at the lower transition region level is smaller than the energy flux required for heating the corona, implying that we may need another heating mechanism. We should, however, note a possibility that the energy dissipated at the chromosphere might be overestimated because of the opacity effect. 
\end{abstract}

\maketitle
\section{Introduction}
Thermal conduction from the solar interior cannot form the solar outer atmosphere, i.e., the chromosphere and the corona, and thus a nonthermal mechanism is required there. Magnetohydrodynamic (MHD) waves have been considered as one of the candidates for the mechanism of the energy transfer to the outer atmosphere. The waves are excited by interactions between magnetic field lines and convective gas flows at the photosphere. They propagate upward along the magnetic field lines, followed by the dissipation of the energy in the upper atmosphere. Depending on the wave modes, frequency, and field topology, a fraction of the waves may reflect back to the lower atmospheric layers. \par
Compressible magnetoacoustic waves may be evolved to shock waves due to steepening, and their dissipation might contribute to the heating of the atmosphere. The temporal profiles of the Doppler velocity measured at the chromosphere show the sawtooth shapes, and they can be a signature of the shock formation \citep{2003A&A...403..277R,2006ApJ...640.1153C,2014ApJ...786..137T}. They also reported intensity enhancements with blue-shifted motion, indicating the strong compression and heating at the shock front. However, magnetoacoustic waves generated at the photosphere are thought to be an insufficient driver to heat the solar corona because of rapid dissipation before reaching the corona \citep{1981A&A....97..310M,1989ApJ...346.1010A}. Therefore, such waves are currently considered as a possible candidate for heating the chromosphere, and we need further quantitative evaluations and their discussions. \par
Alfv$\acute{\mathrm{e}}$n waves are waves in incompressible modes and thus have difficulty evolving to the shock waves and dissipating the energy, compared to the compressible waves. Therefore, they may carry much energy to the corona without dissipating before reaching the corona. {\it Coronal Multi-Channel Polarimeter (CoMP)}, {\it Hinode} \citep{2007SoPh..243....3K} and {\it Atmospheric Imaging Assembly} \citep[{\it AIA}:][]{2012SoPh..275....3P} on board {\it Solar Dynamics Observatory} \citep[{\it SDO}:][]{2012SoPh..275....3P} found that the solar atmosphere is filled with Alfv$\acute{\mathrm{e}}$n waves. \cite{2007Sci...317.1192T} provided the time series of the line-of-sight (LOS) velocity, the intensity, and the linear polarization maps measured with {\it CoMP}, revealing propagating oscillatory signals in large-scaled coronal structures. By using the Solar Optical Telescope \citep[{SOT}:][]{2008SoPh..249..233I,2008SoPh..249..221S,2008SoPh..249..197S,2008SoPh..249..167T} on board {\it Hinode},  \cite{2007Sci...318.1577O} and \cite{2007Sci...318.1574D} found the transverse oscillations in the chromospheric prominences and spicules, suggesting the existence of Alfv$\acute{\mathrm{e}}$n waves. By using the Extreme Ultraviolet Imaging Spectrometer \citep[{EIS}:][]{2007SoPh..243...19C} on board {\it Hinode}, \cite{2012ApJ...753...36H} and \cite{2013ApJ...776...78H} reported the decrease in the nonthermal line widths with heights in the polar coronal hole and suggested a signature for the energy dissipation of Alfv$\acute{\mathrm{e}}$n waves. More recently, the {\it Interface}  {\it Region}  {\it Imaging}  {\it Spectrograph} \citep[{\it IRIS}:][]{2014SoPh..289.2733D} coordinated with the {\it Hinode}/SOT provided a spectroscopic measurement of oscillations in chromospheric prominence threads, suggesting the resonant absorption of Alfv$\acute{\mathrm{e}}$n waves and their subsequent heating \citep{2015ApJ...809...72A,2015ApJ...809...71O}. \par
The MHD waves are dominantly generated by the photospheric motions, and the linkage between the photospheric motions and behaviors at the upper atmosphere is quite important for understanding the heating in the chromosphere and corona. \cite{2009ApJ...692.1211C} studied the MHD waves in the photosphere and the chromosphere by examining the simultaneous photospheric Si I line and chromospheric He I line obtained by the {\it Tenerife Infrared Polarimeter (TIP)} operating at {\it Vacuum Tower Telescope (VTT)}. They reported a variety of the chromospheric oscillations in amplitude, frequency, and stage of shock formation even when quite similar oscillations were observed at the photosphere, implying the importance of the propagating processes related to the magnetic features. \cite{2010ApJ...722..131F} studied the waves in sunspots with He I 10830{\AA}, Ca II H 3969{\AA}, Fe I 3969.3{\AA}, Fe I 3966.6{\AA}, Fe I 3966.1{\AA}, Fe I 3965.4{\AA}, and Si I 10827{\AA}, covering the photosphere and the chromosphere. With the phase difference spectra of LOS velocities between several pairs of lines, they revealed standing waves at frequencies lower than 4 mHz and a continuous propagation of waves at higher frequencies, which is consistent with the slow-mode waves in the stratified atmosphere. Similar results are reported by \cite{2006ApJ...640.1153C} and \cite{2013SoPh..284..379K}. \cite{2011ApJ...735...65F} performed the data-driven MHD simulation of the waves in the sunspot and reported a remarkable agreement with the observations. \par
The connectivity between the photospheric motions and the coronal response is also studied. \cite{2010ApJ...710.1857M} derived the spectrum of the photospheric horizontal velocity from the time series of {\it Hinode}'s G-band images and applied it to their MHD simulation. They found that the Alfv$\acute{\mathrm{e}}$n waves excited by the observed photospheric granular motions can bring enough energy to the corona for the heating. \cite{2005ApJ...621..498K} identified that the footpoints of the hot coronal loops have a lower magnetic filling factor than the footpoints of the cool coronal loops, indicating the importance of the flexibility in the photospheric horizontal motions to heat the corona. \par
An important observational study for understanding the roles of MHD waves in heating the upper atmosphere is to evaluate how much energy the observed MHD waves have at various atmospheric heights. Accurate measurements of physical quantities in the waves are required for the quantitative evaluation. MHD waves can give fluctuations to the magnetic fields, which observers have been attempted to measure with ground-based telescopes \citep{1997AN....318..129L,1998ApJ...497..464L,2000ApJ...534..989B}. These observations, however, may not confidently show that the observed magnetic fluctuations are intrinsic because of the temporal fluctuations of the atmospheric seeing. Observations from space would rather provide more confident results. \cite{2009ApJ...702.1443F} investigated the weak fluctuations in temporal behaviors of spectropolarimetric data from the {\it Hinode}/SOT, suggesting that the phase relations of the photospheric fluctuations in plages and pores can be explained by the dominant existence of standing waves at the photosphere with a small but sufficient leakage toward the chromosphere. It is worth noting that \cite{2009ApJ...702.1443F} used only the photospheric information. \par
\begin{figure}[t]
\centering
\includegraphics[width=15cm]{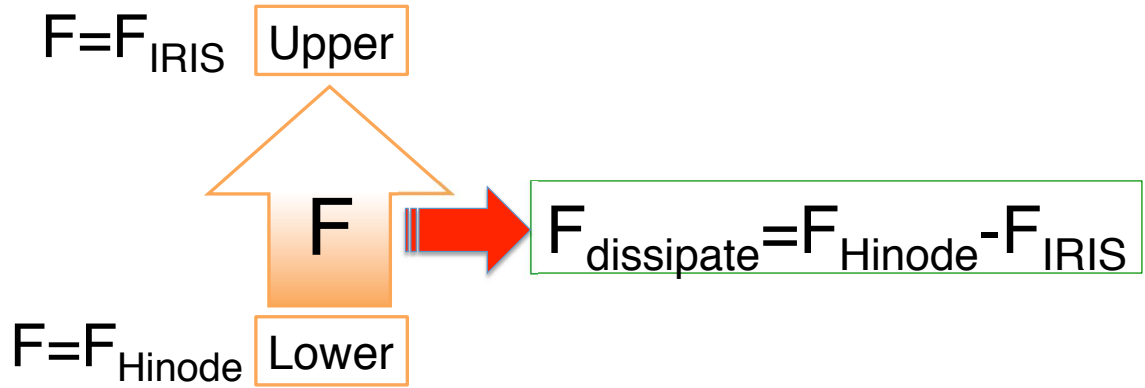}
\caption{Schematic drawing of the logic for estimating the dissipated energy flux}
\label{goal}
\end{figure}
The fundamental motivation of this study is to estimate the dissipated energy of MHD waves in the upper atmosphere with simultaneous multi-height observations. The temporal behaviors of the physical parameters are important for identifying the mode of waves. The time series of the data obtained with ground-based telescopes are less suitable because of the seeing effect. Moreover, rather than imaging observations, spectroscopic observations are preferable for detecting the fluctuations caused by MHD waves in physical quantities quantitatively and accurately. For these reasons, the coordinated {\it Hinode} and {\it IRIS} observations are used in this study. The {\it Hinode}'s spectropolarimetric observations provide the tiny fluctuations in the physical parameters, including the magnetic flux density at the photospheric level, while the {\it IRIS} spectroscopic observations provide the temporal series of intensity and Doppler speeds measured with the chromospheric and the transition region spectral lines. The combination of these observations allows us to trace the temporal behaviors of MHD waves at the two atmospheric layers at the same time. As shown in Figure \ref{goal}, the dissipation rate of the energy can be evaluated with the upward energy fluxes estimated at the two layers. \par
This paper presents a set of {\it Hinode} and {\it IRIS} simultaneous high-cadence observations
and discusses how much energy flux  the MHD waves observed in the data have at the two layers. The time series of the {\it Hinode} data used in the study has a cadence more than two times higher than that used in \cite{2009ApJ...702.1443F}, giving a more valid conclusion of the wave-mode identification. We describe observational methodologies in section 2. Section 3 shows the observational results, which are interpreted and used for the estimate on the energy flux in section 4. A summary of this paper and conclusions are given in section 5.

\section{Observations and data analysis}
{\it Hinode} and {\it IRIS} observed a well-developed leading sunspot of NOAA Active Region 11836 on 2013 September 4. The sunspot was located at ($x$,$y$)=(510\arcsec,75\arcsec) at 16:00 UT in the heliocentric coordinates. In this study, we mainly focus on MHD waves in the sunspot umbra. Since the observed sunspot is not at  the disk center, we divided observed amplitudes by $\cos\theta$, where $\theta$ is a heliolongitudinal angle $\sim$ 31 degree from the meridional line. Here it is assumed that the observed fluctuations are mainly in the direction of the umbral magnetic field, which is almost normal to the solar surface. This assumption will be reasonable according to the mode identification shown later. \par
\begin{figure}
\centering
\includegraphics[width=16cm]{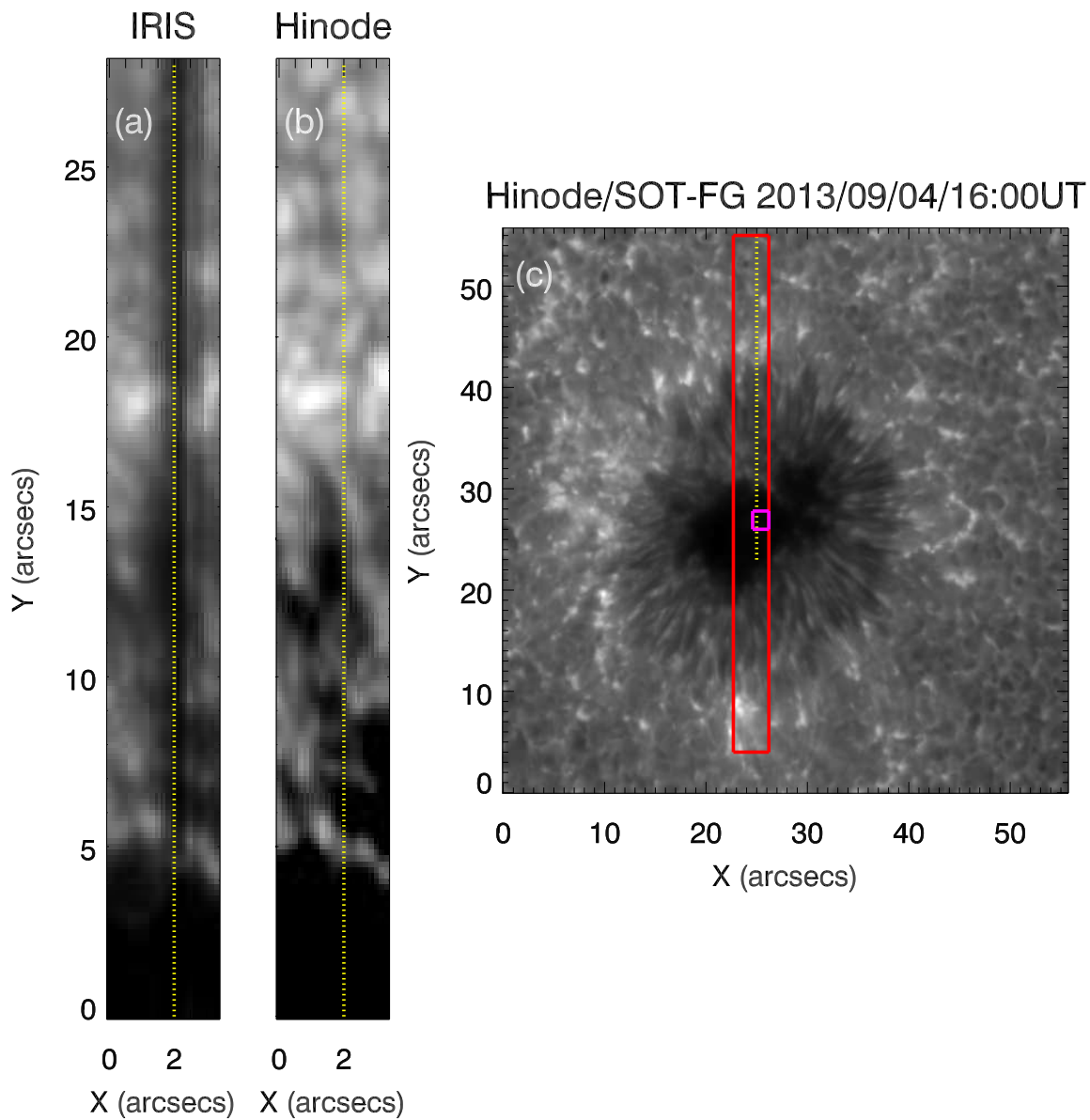}
\caption{(a) An {\it IRIS} slit-jaw image in Mg II wing, co-aligned with (b) a {\it Hinode}/SOT-SP continuum map. Yellow dotted lines are given at the position of the {\it IRIS} slit. (c) A Ca II H image of the sunspot in NOAA Active Region 11836 at 16:00 UT on 2013 September 4, observed with SOT's filtergraph. 
The yellow line gives the slit position for the {\it IRIS} raster data, with the red rectangle giving the field of view of the {\it Hinode}/SOT-SP map. The purple square gives the field of view used in the subsequent sections.}
\label{region}
\end{figure}

Figure \ref{region} is a snapshot of the sunspot observed in Ca II H, with the positional relationship of the data used in this study. The yellow line gives the slit position for the {\it IRIS} raster data, with the red rectangle giving the field of view of the SOT's spectropolarimeter \citep[{SP}:][]{2013SoPh..283..579L} at the observing period. The SP observation was carried out for a region of interest at 15:39-16:31 UT, while the {\it IRIS} observation was carried out at 15:48-17:57 UT. We chose an overlapped part of the observation time (15:48-16:31 UT) for our data analysis.

\subsection{\it Hinode SP observation}
The SP recorded the four Stokes (I, Q, U and V) profiles of the Fe I lines at 6301.5{\AA} and 6302.5{\AA} with a spectral sampling of 21.55 m{\AA}. A 3.8 arcsec range was repeatedly mapped with measurements at 12 slit positions; one spectral measurement with a slit width of 0\arcsec.15 and then the next measurement after moving 0\arcsec.30 in the west direction (a sparse raster scanning). One measurement with the accumulation of photons in 1.6 sec archived the cadence of 27 sec in mapping. The two pixels were summed in the slit direction, providing the spectral data with a pixel size of 0\arcsec.32. We used the calibrated Stokes data (Level 1 data) available via CSAC at HAO/NCAR, which is calibrated with the standard SOT-SP calibration software \citep{2013SoPh..283..601L}.
 
\subsection{\it Hinode data analysis}
For the detection of weak magnetic fluctuations, Stokes V is more preferable to Stokes Q and U because of its much higher sensitivity. We used the Stokes I and V profiles of the Fe I 6301.5{\AA} line to derive the LOS velocity, the LOS magnetic flux density, and the intensity. The LOS velocity was derived by applying a single Gaussian fit to the Stokes I. Since the magnetic filling factor inside the sunspot umbra is almost unity, effect of the nonmagnetic atmosphere is negligible. The intensities at the line core ($I_{core}$) and the continuum ($I_{cont}$) are defined as
\begin{equation}
I_{cont} \equiv \left< \int_{6300.9} ^{6301.0} I(\lambda) d\lambda \right>
\end{equation}
and
\begin{equation}
I_{core} \equiv  \mbox{min} [I(\lambda)] ^{6302} _{6301}.
\end{equation}

\begin{figure}
\centering
\includegraphics[width=16cm]{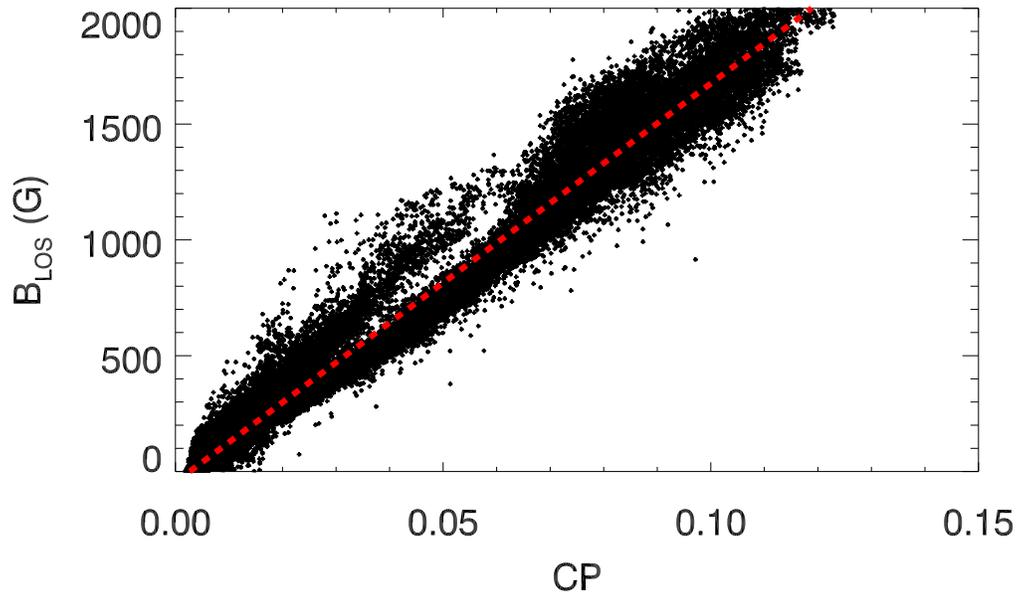}
\caption{Scatter plots of the LOS magnetic flux density and the degree of the circular polarization as defined in equation (\ref{2_a}). The red dashed line indicates a linear regression line.}
\label{fit_regression}
\end{figure}

Following \cite{2009ApJ...702.1443F}, the area of Stokes V profiles was used to derive the LOS magnetic flux density (the so-called 'weak-field approximation'). The weak-field approximation is valid inside sunspot umbrae according to \cite{2014ApJ...795....9F} with synthetic profiles of the Fe I 6301.5{\AA} line. We first calculated the degree of the circular polarization $CP$ as defined by
\begin{equation}
CP=\frac{V}{I_{cont}} \label{2_a},
\end{equation}
where 
\begin{equation}
V \equiv \int_{6301.0} ^{6302.0} |V|(\lambda) d\lambda.
\end{equation}
A coefficient is needed to convert the $CP$ to the LOS magnetic flux density $B_{LOS}$. The coefficient was determined by a linear regression line in the scatter plot between the $CP$ and $B_{LOS}$ derived from a Milne-Eddington inversion (Figure \ref{fit_regression}). The linear regression is given by
\begin{equation}
CP =(5.8 \times 10^{-5})B_{LOS}+0.0027 \label{coeff}.
\end{equation}
The data used here are all the SP spectra taken during 15:43-15:56 UT. \par
It should be noted that the Stokes inversion with the Milne-Eddington atmosphere may be subject to the photon noise, which impedes the detection of weak fluctuations in magnetic flux density because the inversion needs to determine a lot of free parameters. In addition, since the Milne-Eddington inversion can fit only symmetric Stokes profiles, slight asymmetric shapes of the observed Stokes profiles also impede the detection of weak fluctuations. Actually, even inside sunspot umbrae where the asymmetry in Stokes profiles is relatively small, there is about 20 G standard error because of the asymmetry \citep{2010ApJ...720.1281G}. When the observed magnetic fluctuation profiles inside the sunspot umbra were derived with the Milne-Eddington inversion, they were noisy. Note that such errors do not affect to the coefficient of proportionality in equation (\ref{coeff}) because the errors are sufficiently smaller than the background (not fluctuated) magnetic fields.
\subsection{\it IRIS observation}
{\it IRIS} performed spectroscopic observations of the chromosphere and the transition region. The UV spectral data were acquired with sit-and-stare mode; the slit (0\arcsec.33 width) was pointed at one solar location, and its position was in the SOT-SP's field of view. The acquired spectral data cover two wavelength ranges; near-ultraviolet (NUV) range including Mg II at 2796{\AA} and 2803{\AA} (10$^{4.0}$ K) and far-ultraviolet (FUV) range including Si IV at 1403{\AA} (10$^{4.8}$ K) and O IV at 1400, 1401, and 1405{\AA} (10$^{5.2}$ K), where the value in parentheses is the formation temperature of each line. The spectral resolutions of the NUV and FUV ranges are 12.72 m{\AA} and 25.46 m{\AA}, respectively. The pixel size along the slit direction is 0\arcsec.17 and the cadence is 3 seconds. At the same time, the series of the slit-jaw images (SJIs) for Si IV ($10^{4.8}$ K), C II ($10^{4.2}$ K), Mg II ($10^{4.0}$ K) and Mg II wing ($10^{3.7}$ K) were obtained every 12 seconds. Their field of view is 35\arcsec$\times$40\arcsec and they were used to identify the exact location of the slit on the solar features. We used the level 2 data created with the instrumental calibration including the dark current subtraction, flat field, and geometrical corrections \citep{2014SoPh..289.2733D}.

\subsection{\it IRIS data analysis}
We applied a single gaussian fit to the Mg II k 2796{\AA} and Si IV 1403{\AA} spectra independently to derive the LOS velocity at two different temperatures. Here the center position of each spectral line averaged over the field of view was used as the reference wavelength. Since the Mg II spectral line, which has large opacity, is formed in a non-local thermodynamic equilibrium condition, a central reversal is typically observed in the line core \citep{2013ApJ...772...89L,2013ApJ...772...90L,2013ApJ...778..143P}. However, note that the Mg II lines observed in sunspot umbrae have no central reversed profiles as reported by \cite{2001ApJ...557..854M}. The intercombination multiplet of O IV lines at 1397.2, 1399.8, 1401.2, 1404.8 and 1407.4{\AA} provides a well-known set of density-sensitive pairs. We used the ratio of the 1399.8{\AA} and 1401.2{\AA} lines for electron density, and we did not use the other lines because of our spectral coverage and line blending \citep{2015arXiv150905011Y}.

\subsection{Data co-alignment}
The time series of the SP mapping data was aligned spatially by performing the local cross-correlation of the SP continuum image with the subsequent frame. The time series of the SJIs at the Mg II wing was also aligned with the same procedure. In the both alignments, photospheric sunspot features such as umbrae and penumbrae worked as fiducial marks. Then, the SP maps were co-aligned with the {\it IRIS} images by using the SP continuum image and the Mg II wing SJI at the start of each time series. Bright features seen in outside the sunspot were used as fiducial marks for the SP-{\it IRIS} co-alignment. The aligned field of view is shown in Figure {\ref{region}} (a)(b). The IDL procedure {\tt{get${\_}$correl${\_}$offsets.pro}} was used to get a rigid displacement in the cross-correlation. Note that SP maps were stretched in the X-direction before the co-alignment because of the sparse raster mapping. In addition, the pixel scale of the SP maps was scaled to that of the {\it IRIS} SJIs by using the IDL procedure {\tt{congrid.pro}}. The accuracy of the co-alignment is better than 0\arcsec.5 according to the visual inspection of the co-aligned data. The slit position seen in the time series of SJIs was checked to confirm the positional fluctuations of the slit on the solar surface with a magnitude of much less than 1\arcsec. 
\section{Results}
\subsection{Oscillations at the photosphere observed with {\it Hinode}
  \label{sec:results:Hinode}}
\begin{figure*}
\centering
\includegraphics[width=16cm]{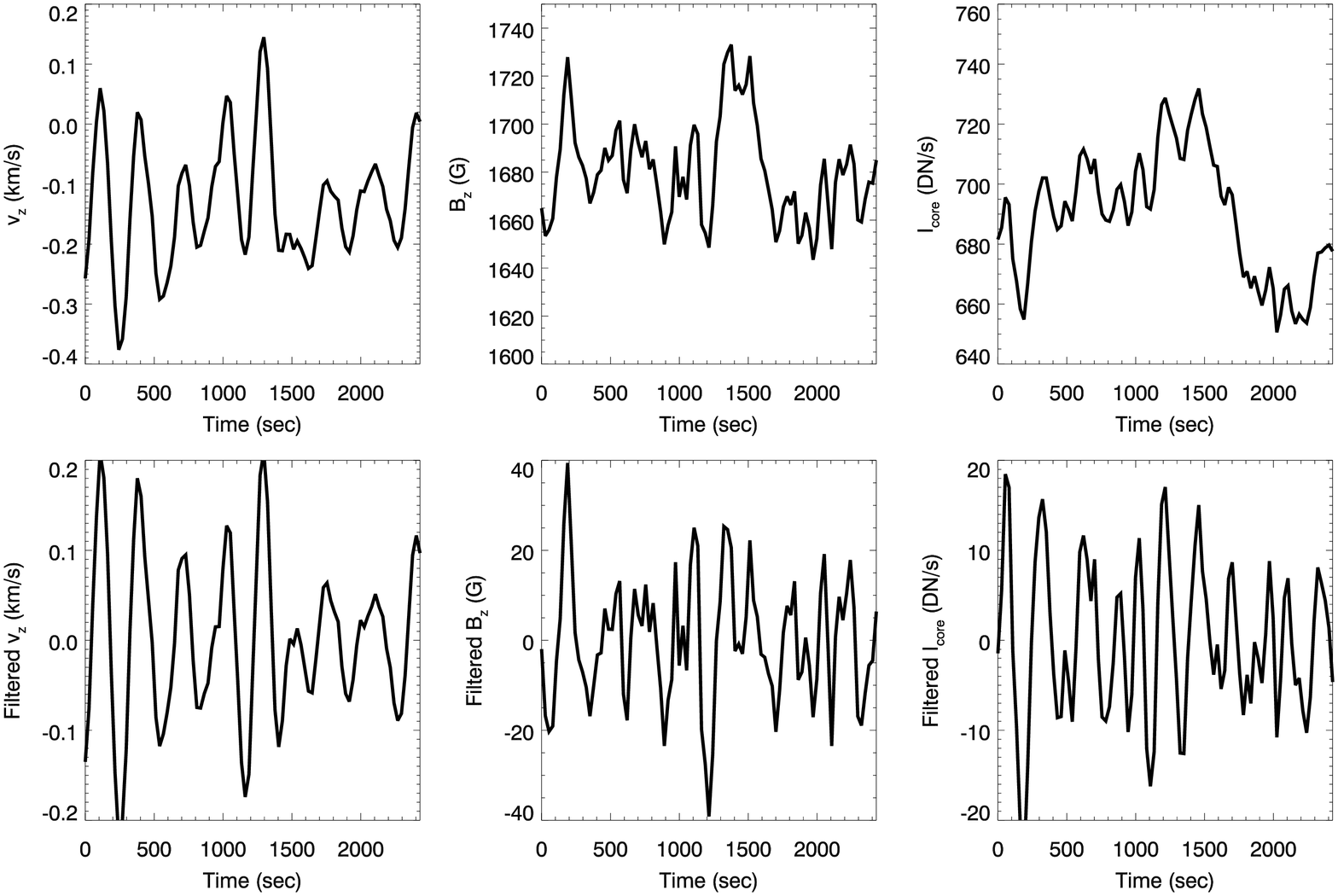}
\caption{From the left to the right, time series of the Doppler velocity, the magnetic flux density, and the intensity in the line core, derived from the SP Fe I 6301.5{\AA} measurements. The lower panels are their residuals ($\delta v_z$, $\delta B_z$, $\delta I_{core}$) after subtracting the 12 points running average from the original time series. Positive and negative velocities imply blueshift and redshift, respectively.}
\label{running_average}
\end{figure*}
\begin{figure}
\centering
\includegraphics[width=16cm]{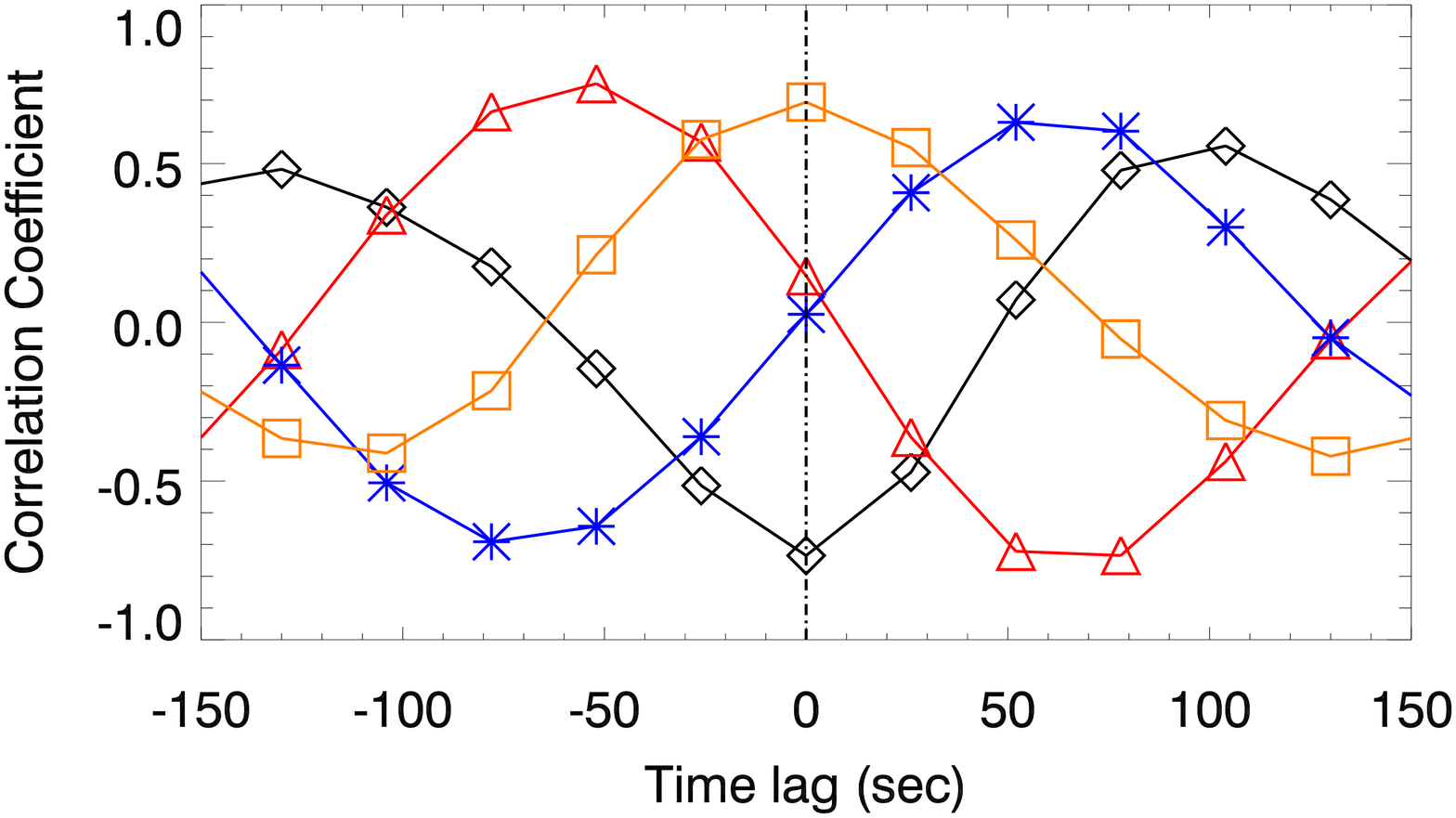}
\caption{Correlation coefficient for physical parameters observed in the sunspot umbra as a function of time lag. Each symbol implies the pair of physical parameters. Black diamond, blue asterisk, red triangle and orange square show correlation coefficients on $\delta I_{core} - \delta B_z$, $\delta v_z-\delta I_{core}$, $\delta v_z- \delta B_z$ and $\delta I_{core}- \delta I_{cont}$ respectively. Note that the correlation coefficient was obtained for each time lag by giving the time lag to the time profile of the latter in the two physical parameters.}
\label{cc_sp}
\end{figure}
Figure \ref{running_average} shows the temporal evolution of the Doppler velocity, the magnetic flux density, and the line core intensity, derived from the SP data averaged in the 6$\times$6 pixel (1\arcsec.92$\times$1\arcsec.80) area inside the sunspot umbra specified by the purple square in Figure \ref{region} (c). Periodic oscillating features are visible in the profiles. The dominant periods are around 5 minutes. Note that we subtracted the 12 points (324 sec) running average from the original time series data to remove the long-term gradual change in the profiles. Since the wave features are similar to sinusoidal functions, we derived the amplitude of the fluctuations by multiplying $\sqrt{2}$ by the root-mean-square values. The results are tabulated in Table 1, where the subscripts z and 0 means that these values are perpendicular components to the solar surface and absolute values, respectively. The typical scale factor for DN is about 76 charges in a CCD pixel \citep{2013SoPh..283..579L}. The uncertainties in $\delta B_z$, $\delta v_z$, $\delta I_{core}$ and  $\delta I_{cont}$ were estimated to be 2.9 G, 0.0014 km s$^{-1}$, 0.37 DN s$^{-1}$ and 0.60 DN s$^{-1}$, respectively. The uncertainties in $\delta B_z$ and $\delta v_z$ derived with Stokes V profiles were obtained by taking account of the standard deviation of intensity fluctuations by photon noise in a continuum range of Stokes V profiles. The uncertainties in $\delta I_{core}$ and $\delta I_{cont}$ derived with Stokes I profiles were estimated to be $\sqrt{\mbox{(photon count})}$ assuming the Poisson distribution. \par

\begin{figure}
\centering
\includegraphics[width=9cm]{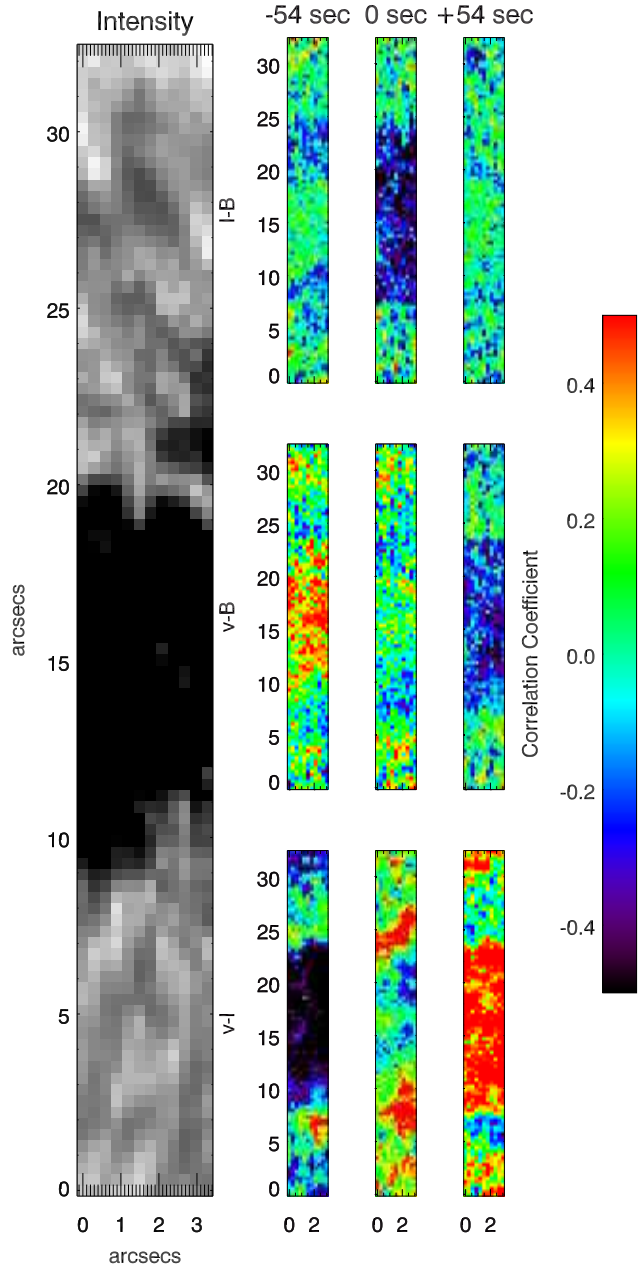}
\caption{Spatial distribution of the cross-correlation coefficients among the physical parameters with three time lags (-54 sec, 0 sec and +54 sec). The sunspot umbra is located at the center and surrounded by the penumbra in the field of view, as shown in the intensity image. Right, from top to bottom: spatial distribution of the cross-correlation coefficient on $\delta I_{core} - \delta B_z$, $\delta v_z-\delta B_z$, and $\delta v_z - \delta I_{core}$, respectively.}
\label{phase_map}
\end{figure}

\begin{table*}[t]
\begin{center}
\begin{tabular}{|c|c|c|c|c|c|c|} \hline
$\delta B_{z}$ & $\delta v_z$ & $\delta I_{core}$ & $\delta I_{cont}$ & $\delta B_z/B_0$ & $\delta I_{core}/I_{cont}$ & $\delta I_{cont}/I_{cont}$ \\
(G) & (km s$^{-1}$) & (DN s$^{-1}$) & (DN s$^{-1}$) &(\%) &(\%) &(\%) \\ \hline 
20 $\pm$ 2.9& 0.13 $\pm$ 0.0014& 11 $\pm$ 0.37 & 12 $\pm$ 0.60 & 0.98 $\pm$ 0.15 & 0.70 $\pm$ 0.015 & 0.78 $\pm$ 0.024 \\ \hline
\end{tabular}
\label{parameters}
\end{center}
\caption{Amplitude of the oscillations observed at the photosphere in the sunspot umbra with {\it Hinode}/SOT-SP.}
\end{table*}
To determine the phase relations among the Doppler velocity, the magnetic flux density, the core intensity, and the continuum intensity, we obtained cross-correlation coefficients in the time profile between two quantities from these four parameters. Figure {\ref{cc_sp}} shows the cross-correlation coefficients between two of the observed parameters as a function of the time lag. The cross-correlation coefficient was obtained when a time lag was given to one of the two time profiles. Such calculations were made for 11 different time lags. The correlation coefficient between $\delta I_{core}$ and $\delta I_{cont}$ is at maximum with no time lag, meaning that there is no phase shift between $\delta I_{core}$ and $\delta  I_{cont}$. The correlation coefficient between $\delta I_{core}$ and $\delta B_z$ is at minimum with no time lag, implying a phase difference between $\delta I_{core}$ and $\delta B_z$ by the $\pi$ radians (180$^\circ$). The correlation coefficient between $\delta v_{z}$ and $\delta B_{z}$ is close to zero with no time lag and gradually decreases with increasing the time lag, meaning that the $\delta v_{z}$  time profile is delayed by $\frac{\pi}{2}$ radians (90$^\circ$) from the $\delta B_{z}$ time profile. Similarly, the $\delta v_z$ is by $\frac{\pi}{2}$ radians (90$^\circ$) ahead of $\delta I_{core}$. Since the phase relations among the magnetic flux density, the Doppler velocity, and the core intensity depend on wave mode \citep{2009ApJ...702.1443F}, the phase relations among these values are important for identifying the mode of the observed waves, which will be discussed in section \ref{sec:discussion:Mode identification:Photospheric}. The phase relations described above are common at any locations inside the sunspot umbra, as shown in Figure \ref{phase_map}, which shows the spatial distribution of the cross-correlation coefficients of the physical parameters at three time lags (-54, 0, and +54 secs). Note that the time lag of 54 sec corresponds to
$\frac{\pi}{4}$ of the oscillation.
\subsection{Oscillations at the chromosphere and the lower transition region observed with {\it IRIS}
\label{sec:results:IRIS}}
\begin{figure*}
\centering
\includegraphics[width=16cm]{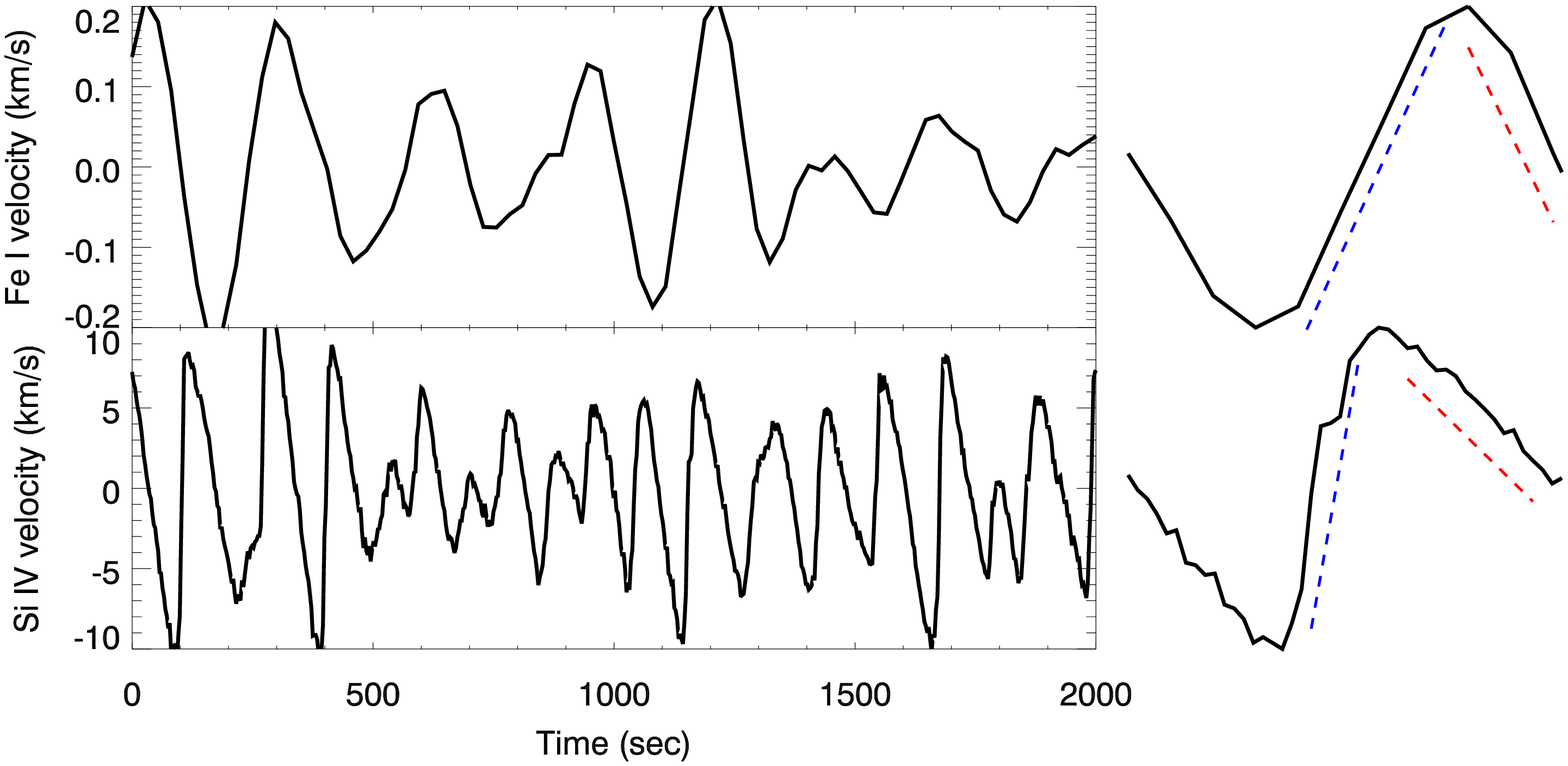}
\caption{Temporal evolution of the Doppler velocities derived from the photospheric Fe I  line at 6301.5{\AA} and the Si IV lower transition region line. Positive and negative values are blueshift and redshift, respectively. An enlarged view on one period waveform is plotted on the right. Blue and red dashed lines show the slope of ascending and descending velocity profiles in the waveform, respectively.}
\label{Hinode_IRIS}
\end{figure*}
\begin{figure}
\centering
\includegraphics[width=16cm]{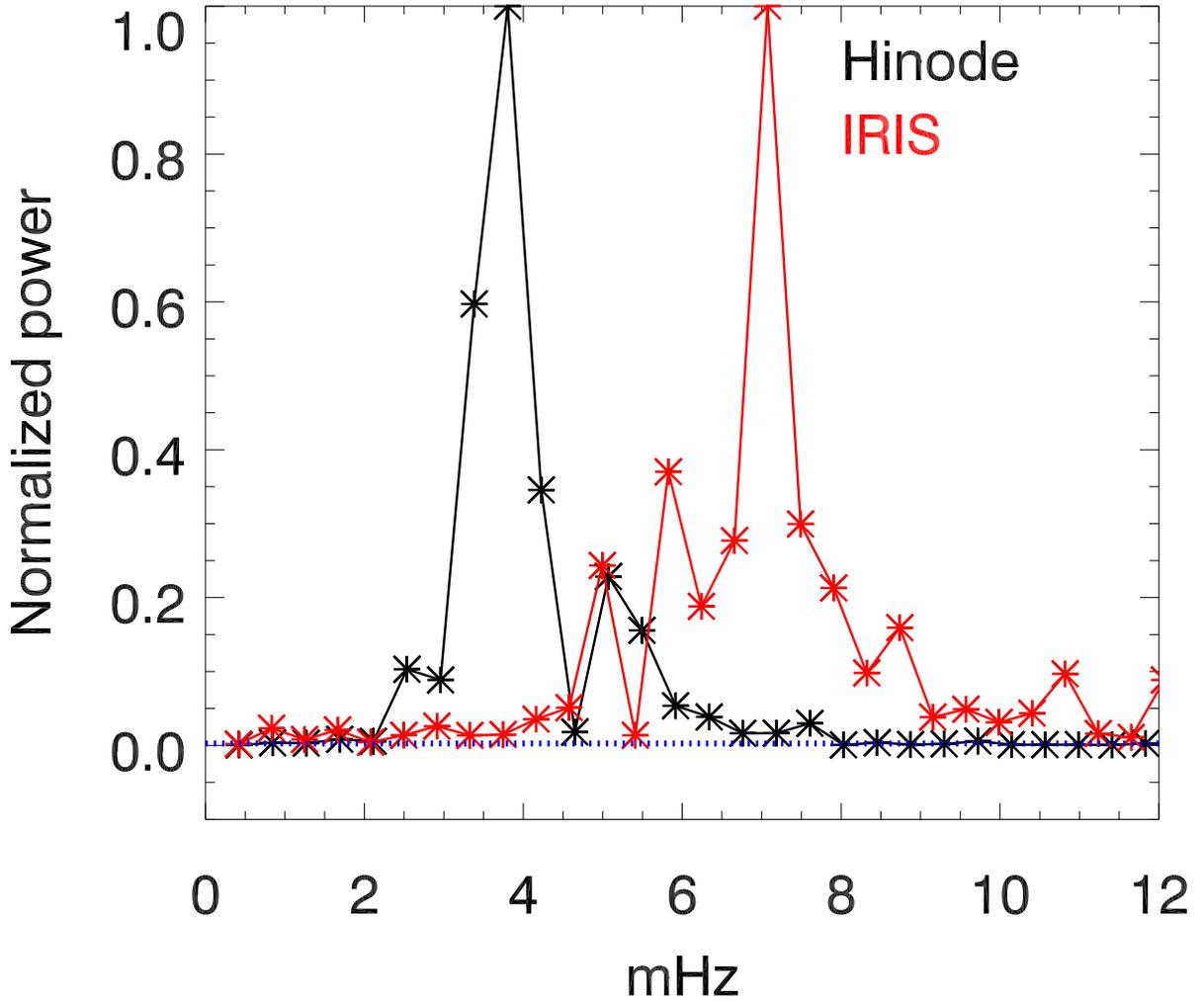}
\caption{Normalized power spectra of the Doppler velocities measured in the sunspot umbra. The black line gives the power spectrum for the Doppler velocities measured with the photospheric Fe I line, whereas the red line gives that of the Si IV line. The blue dotted line is a noise level for the photospheric power spectrum, which is calculated from the average in $>$10 mHz.}
\label{photo_power}
\end{figure}
\begin{figure}
\centering
\includegraphics[width=16cm]{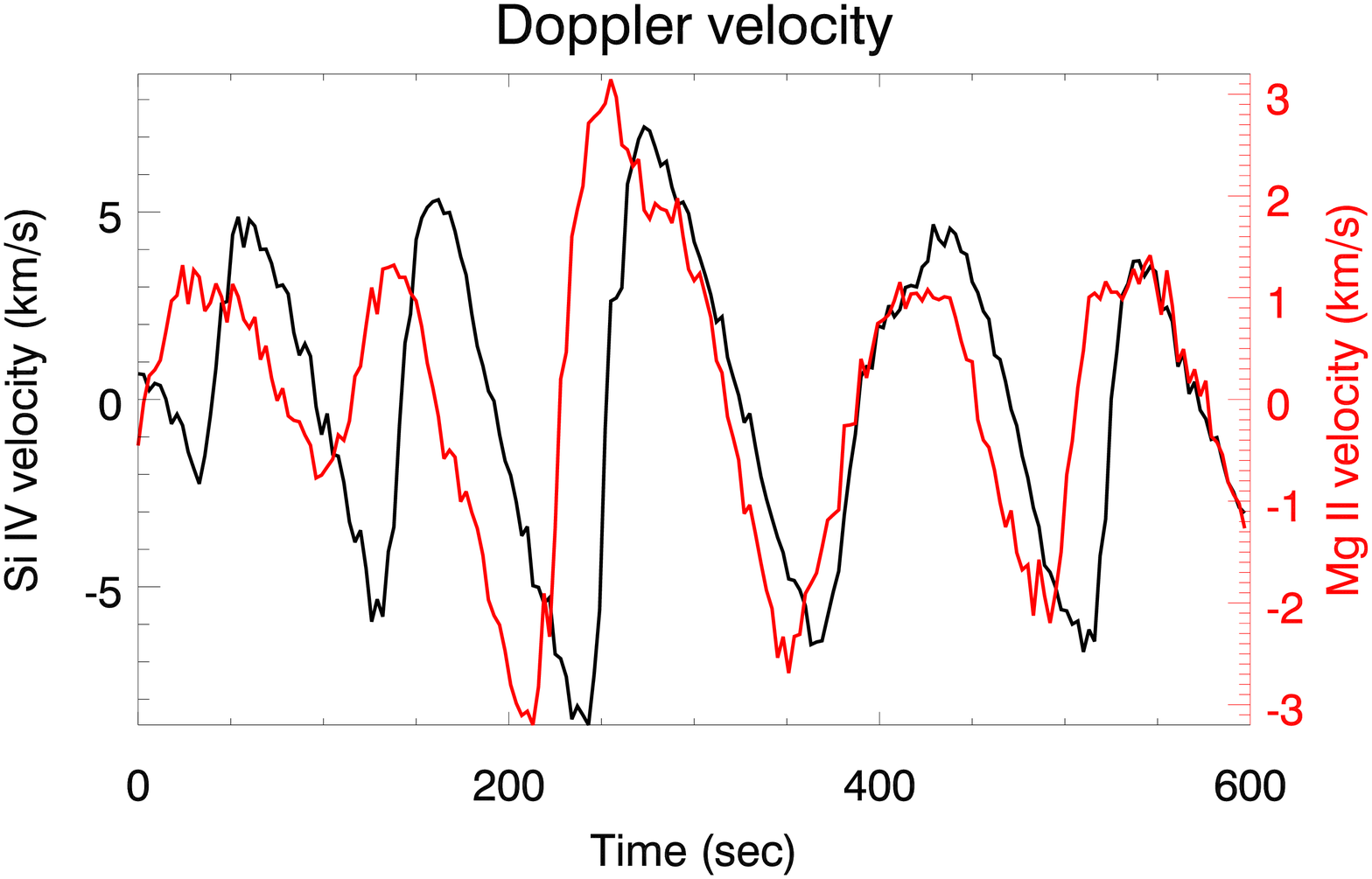}
\caption{Temporal evolution of the LOS velocity at the sunspot umbra. The LOS velocities derived from Mg II k and Si IV are plotted by red and black lines, respectively.}
\label{over}
\end{figure}

Figure \ref{Hinode_IRIS} shows the temporal evolution of the Doppler shift measured with the photospheric Fe I 6301.5{\AA} line, compared with the corresponding profile of the Si IV lower transition region line. The Doppler velocity of the Si IV line is derived from the spectral profile averaged in 3 pixels along the {\it IRIS} slit, which is overlapped with the region of interest given by the purple square in Figure {\ref{region}}. Compared to the Fe I profile, the sawtooth pattern is clearly seen in the temporal evolution of the Si IV velocity. The waveforms observed in the Si IV profile have higher frequency than those in the Fe I profile. The same nature of the waves can be seen in the Fourier power of the velocities, as shown in Figure {\ref{photo_power}}. We subtracted the 324 sec running average from the both original profiles before calculating the Fourier transform. Therefore, the orbital effect of the satellite (about 90 minute cycle) is negligible. Figure {\ref{over}} shows the temporal evolution of Doppler velocities measured with the chromospheric Mg II k line and Si IV lower transition region line. The oscillation in the Si IV time profile is about 20 sec delayed from the oscillation in the Mg II k profile. Note that a similar behavior can be found in \cite{2014ApJ...786..137T}. The amplitude of the Mg II k and Si IV oscillations is 2.0 km s$^{-1}$ and 6.2 km s$^{-1}$, respectively. The electron density ($N_{e}$) was derived by using a pair of emission lines (O IV 1399.8{\AA}/1401.2{\AA}). With CHIANTI v8.1 \citep{2015A&A...582A..56D}, it is $2.6\times10^{10} $ cm$^{-3}$.

\section{Discussions}
  \label{chap:discussion}
In this section, we estimate the dissipated energy flux at the chromosphere. For estimating the energy flux, we need to identify the wave mode (Section \ref{sec:discussion:Mode identification}). After the mode identifications, we will estimate the energy fluxes at both  the photosphere and the lower transition region with the observed amplitudes (Section \ref{sec:discussion:Energy estimation}). Comparing the energy at the photosphere and the lower transition region, we discuss the dissipated energy of the observed MHD waves and its implications for the heating of the solar atmosphere (Section \ref{sec:discussion:Implications}).

\subsection{Mode identification of the waves
\label{sec:discussion:Mode identification}}
For mode identifications, we use the following observed results:
\begin{itemize}
\item The phase relations between two of the intensity $\delta I_{core}$, the magnetic flux density ${\delta B_z}$, and the Doppler velocity ${\delta v_z}$ are determined: $\pi$ radians in $\delta I_{core}$ - $\delta B_z$, -$\frac{\pi}{2}$ radians in $\delta v_z$ - $\delta B_z$, and $\frac{\pi}{2}$ in $\delta v_z$ - $\delta I_{core}$. 
\item The dominant frequency of the chromospheric waves is higher than that of the photospheric waves.
\item The wave oscillation in the lower transition region Si IV line is about 20 sec delayed from that in the chromospheric Mg II k line.
\end{itemize}

\subsubsection{The wave mode at the photosphere
\label{sec:discussion:Mode identification:Photospheric}}
The appearance of fluctuations in the temporal evolution of the intensity can rule out the incompressible mode, because the intensity fluctuation is proportional to the fluctuation of the electron density even in the optically thick condition. In the MHD theory, there are two compressible wave modes, i.e., fast mode and slow mode. The difference between the fast-mode and slow-mode waves is the phase relation of restoring forces. For the fast-mode waves, the phase relation between the gas pressure and magnetic pressure is in-phase. It becomes the opposite for the slow-mode waves, i.e., the out-phase relation between the gas pressure and magnetic pressure. The gas pressure and magnetic pressure are proportional to the intensity and  magnetic flux density, respectively. Thus, for the fast-mode waves, there is no phase difference in temporal evolution between the magnetic flux density and the intensity, whereas the phase difference is $\pi$ radians for the slow-mode waves. Our observations show that the phase difference is close to $\pi$ radians. Thus, we can rule out the fast-mode waves. For slow-mode waves, according to \cite{2009ApJ...702.1443F}, the observed phase relations, i.e., $\pi$ radians between the intensity and magnetic flux density, -$\frac{\pi}{2}$ between the Doppler velocity and magnetic flux density, and $\frac{\pi}{2}$ between the Doppler velocity and intensity, suggest the dominant presence of standing waves. For the above reasons, we suggest that standing slow-mode waves are dominant at the photosphere. 

\subsubsection{The wave mode at the chromosphere and the lower transition region
\label{sec:discussion:Mode identification:Chromospheric}}
Since {\it IRIS} cannot perform spectropolarimetric observations, we cannot identify the wave mode by using the phase relations of the observed parameters. On the other hand, {\it IRIS} observes not only one line but several lines. Considering the different formation heights of the chromospheric Mg II k line and the lower transition region Si IV line, the clear phase difference in these lines shown in Figure {\ref{over}} implies that the chromospheric waves propagate upward. The observed time lag between Mg II k and Si IV is around 20 second. The difference of the height in the line formation between Mg II k and Si IV is about 0.5 Mm \citep{2015ApJ...811...81R}, and thus their propagating speed is roughly 25 km s$^{-1}$, which is close to the sound speed in the atmosphere where Mg II k (T$\sim$10000 K and $c_s \sim$ 15 km s$^{-1}$) and Si IV (T$\sim$80000 K and $c_s \sim$ 40 km s$^{-1}$) are formed. A steepening is observed with {\it IRIS} as a possible sign of shock formation and energy dissipation. Since longitudinal waves are easily steepened compared to transverse waves, the observed steepening signature also supports the identified slow-mode waves at the photosphere. The dominant frequency of the chromospheric waves is $\sim$ 7 mHz, whereas the observed dominant frequency is $\sim 3.7$ mHz at the photosphere. The similar high-frequency enhancements were reported by \cite{2006ApJ...640.1153C,2009ApJ...692.1211C} in the sunspot umbra. The change of the dominant power to higher frequency can be explained with the acoustic cutoff. The oscillations below the cutoff frequency do not propagate upward. On the other hand, above the cutoff value, waves propagate upward freely into the chromosphere. Photospheric standing mode is a consequence of cut and reflected waves, 
because the frequencies of almost all the photospheric waves are below the cutoff frequency, which is roughly $\sim$ 6 mHz, i.e. the lower edge of the strong {\it IRIS} power (Figure \ref{photo_power}).

\subsection{Energy estimation                                                                         
\label{sec:discussion:Energy estimation}}
In this section, we estimate the energy flux based on the identified wave mode (dominant photospheric standing slow-mode waves with leakages of the high-frequency wave components to the chromosphere) and the observed amplitudes. \par
The energy flux ${\bf F}$ is generally written by
\begin{equation}
{\bf F}=\rho \delta v^2 {\bf v}_g + (\delta {\bf v}\times{\bf B})\times \delta {\bf B} \label{flux1},
\end{equation}
where $\rho$, ${\bf B}$, $v$, and ${\bf v}_g$ are the mass density, the magnetic field strength, the velocity amplitude, and the group velocity, respectively. The first and second terms on the right-hand side are thermal-kinetic energy flux and Poynting flux, respectively. The energy flux of the slow-mode wave is described as
\begin{equation}
{\bf F}=\rho \delta v^2 {\bf v_g}.
\end{equation}
Note that since the direction of $\delta {\bf v}$ is the same as {\bf B} in the case of slow-mode waves, the Poynting flux term,
\begin{equation}
(\delta {\bf v}\times{\bf B})\times \delta {\bf B},
\end{equation}
is zero. 
\subsubsection{Energy flux at the photosphere
\label{sec:discussion:Energy estimation:Photospheric}}
For estimating the energy flux, we need to estimate the mass density at the photospheric height. Assuming a uniform straight cylinder as a flux-tube model, \cite{2013A&A...551A.137M} analytically calculated that the photospheric phase speed for the slow-mode waves can be written by
\begin{equation}
\frac{\omega}{k}=c_s \sqrt{\frac{\delta I_{cont}/I_{cont}}{\delta B_z/B_0}} \left[\frac{2}{3}\frac{h\nu}{k_B T} +\frac{\delta I_{cont}/I_{cont}}{\delta B_z/B_0} \right]^{-1/2}. \label{4_e}
\end{equation}
The phase speed of slow-mode waves is close to $c_T$ \citep{1982SoPh...76..239E}, where the tube speed $c_T=\frac{c_s v_A}{\sqrt{c_s^2 +v_A ^2}}$, the sound speed $c_s=\sqrt{\frac{\gamma k_B T}{m}}$ and the Alfv$\acute{\mathrm{e}}$n speed $v_A =\frac{B_0}{\sqrt{4\pi \rho}}$. Therefore, the comparison between equation (\ref{4_e}) and $c_T$ gives $\rho=5.0\times10^{-6}$ g cm$^{-3}$ by substituting the observed parameters (Table 1) and $T=4500$ K. Figure \ref{photo_power} suggests that the waves with the frequency above 6 mHz can penetrate into the chromosphere. Thus, the upward energy flux at the photosphere ($F_{Hinode}$) is estimated by using the Doppler velocity amplitude $\delta v_z =0.027$ km s$^{-1}$, which is derived from the 6-10 mHz data and is sufficiently larger than the noise level estimated by photon noise (0.0014 km s$^{-1}$). Note that the strong {\it IRIS} power exists in the 6-10 mHz range. The waves in the 6-10 mHz may propagate to the chromosphere because of a frequency higher than the cutoff frequency. With $\rho=5.0\times10^{-6}$ g cm$^{-3}$ and $|{\bf v_g}|=c_s=5.4$ km s$^{-1}$, we derive $F_{Hinode}$=$2.0 \times10^{7}$ erg cm$^{-2}$ s$^{-1}$.

\subsubsection{Energy flux at the lower transition region
\label{sec:discussion:Energy estimation:Chromospheric}}
The energy flux of the waves at the formation height of the Si IV line is estimated with the observed amplitude of the Doppler velocity, i.e., $\delta v_z=6.2$ km s$^{-1}$. Here we use the sound speed of $40$ km s$^{-1}$, calculated with the formation temperature of Si IV and the mass density $\rho$ of $5.4\times10^{-14}$ g cm$^{-3}$. The mass density is given by $\rho =N_e \mu m_p$, where $m_p$ is the proton mass ($m_p=1.67\times10^{-24}$ g) and $\mu=1.25$ from the  solar atomic abundance $H:He=3:1$. The electron density ($N_e$) used here is $2.6\times10^{10} $ cm$^{-3}$, which was derived from a pair of emission lines (O IV 1399.8{\AA}/1401.2{\AA}). Note that the plasma observed with the O IV lines is almost the same as that what with Si IV, as reported by \cite{2016ApJ...817...46M}. With these parameters, we obtained an energy flux of $8.3\times10^4$ erg cm$^{-2}$ s$^{-1}$. The corona above sunspot umbrae is sometimes dark in soft X-rays. However, \cite{2000ApJS..130..485N} reported that sunspot temperatures  and emission measures at the corona are still lower than the average active region parameters but higher than the quiet region plasma parameters. Since the coronal energy loss at the quiet region is about $ 3\times10^5$ erg cm$^{-2}$ s$^{-1}$ \citep{1977ARA&A..15..363W}, which is larger than our estimated energy flux at the lower transition region, we can say that our estimated energy flux is not enough for the requirement of the coronal heating. Furthermore, we should note that the estimated density might be overestimated by up to several factors, because of the nonequilibrium ionization effect \citep{2013ApJ...767...43O,2015arXiv150905011Y}. Since the density is proportional to the energy flux, the energy flux might also be overestimated by up to several factors.\par

\subsection{Implications for the heating of the solar atmosphere
\label{sec:discussion:Implications}}
\begin{figure}
\centering
\includegraphics[width=16cm]{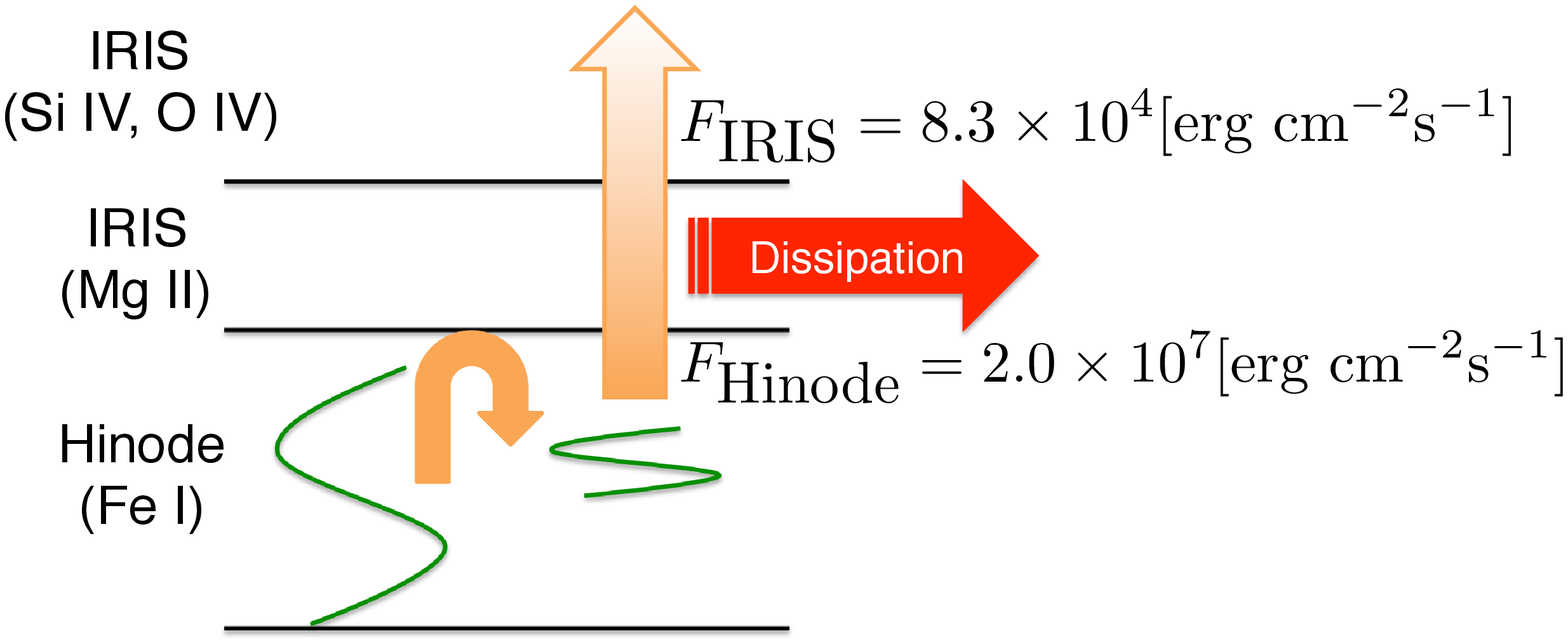}
\caption{Summary of the energy fluxes estimated in this study}
\label{schematic}
\end{figure}
The energy fluxes estimated in this study are summarized in Figure \ref{schematic}. The difference of the energy flux between $F_{Hinode}$ and $F_{IRIS}$ may be considered as the amount of the energy dissipated by the waves before they reach the transition region level. The dissipated energy flux is enough to heat the umbral chromosphere (about $2\times10^6$ erg cm$^{-2}$ s$^{-1}$ from \cite{1981phss.conf..235A} and \cite{1985JKAS...18...15L}). It means that the dissipation of the compressible shock waves is crucial to form the umbral chromosphere. Since the magnetic field in sunspot umbrae is highly bundled, we guess that the discontinuity of the magnetic field is not likely to be created inside umbral fields. Therefore, small energy releases such as nanoflares might not contribute to the atmospheric heating in sunspot umbrae. The energy flux observed with the Si IV line is much smaller than the energy input required for the coronal heating in umbrae. This suggests that other heating mechanisms may be important in the corona, at least in the coronal magnetic structures connecting to sunspot umbrae. \par

\begin{figure}
\centering
\includegraphics[width=16cm]{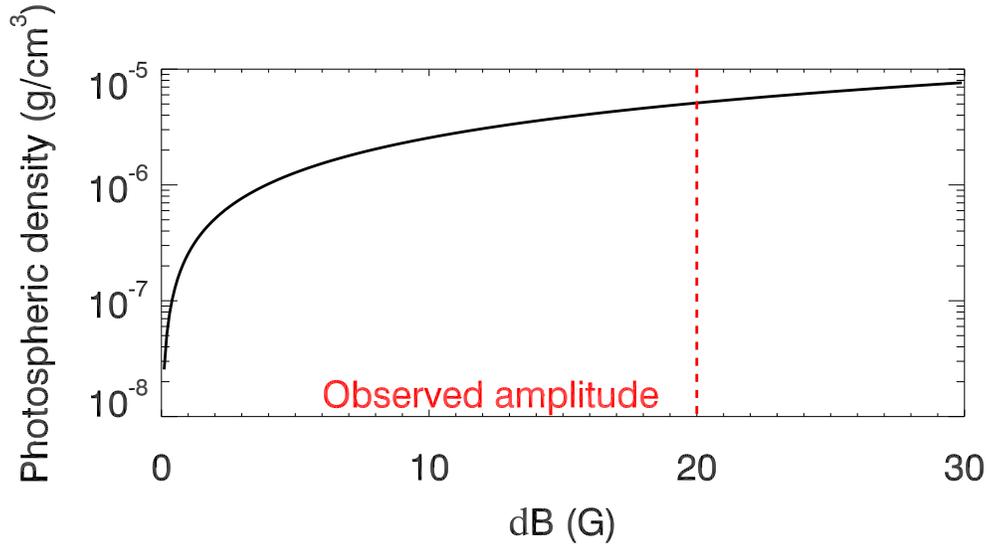}
\caption{Photospheric mass density derived with the seismology discussed in section \ref{sec:discussion:Energy estimation:Photospheric} as a function of intrinsic fluctuations in magnetic flux density. The red dashed line gives the observed amplitude of the fluctuations. Note that there is a possibility that a part of the observed fluctuations is not an intrinsic fluctuation in the magnetic flux density.}
\label{dB_dependence}
\end{figure}

We should note that our estimated photospheric density is larger than that in standard empirical atmospheric models, such as \cite{1986ApJ...306..284M} and \cite{2006ApJ...639..441F}. As an example, in the \cite{1986ApJ...306..284M}'s model, the mass density $\rho$ is less than $10^{-7}$ g cm$^{-3}$ at $z=300$ km which corresponds to the formation height of the Fe I 6301.5{\AA} line \citep{2014ApJ...795....9F}. If we assume the photospheric density with \cite{1986ApJ...306..284M} model, the dissipated energy flux becomes smaller than the requirement for the chromospheric heating. Therefore, it is quite important to understand the reasons of the discrepancy. We have following three ideas. \par
The opacity effect may be one of the reason why the photospheric density estimated with the seismology is relatively a large value, as discussed in \cite{1998ApJ...497..464L}, \cite{2000ApJ...534..989B}, \cite{2003ApJ...588..606K} and \cite{2014ApJ...795....9F}. Temperature and density fluctuations associated with the propagation of compressible waves may cause fluctuations in opacity; The line formation layer is moved upward and downward, resulting in an apparent fluctuation in magnetic flux density. For estimating the photospheric density, we assume here that the observed fluctuations of magnetic flux density are fully intrinsic ($\delta B=\delta B_{intrinsic}$). However, there is a possibility that the opacity change may cause a false signal in the fluctuations in the magnetic flux density ($\delta B=\delta B_{intrinsic}+\delta B_{opacity}$). There is no phase difference between the density increase and the rising motion of the line formation height. Thus when we only consider the opacity effect caused by density fluctuation, the phase difference between $\delta I_{core}$ and $\delta B_z$ is observed as out of phase ($\pi$ radians), which is same as what we observed. This means that the observed $\delta B$ gives the maximum value of $\delta B_{intrinsic}$. Figure \ref{dB_dependence} shows the photospheric mass density derived by the seismology as a function of $\delta B_{intrinsic}$. This shows that the density becomes small when $\delta B_{intrinsic}$ becomes small. By performing a numerical simulation, \cite{2003A&A...410.1023R} suggested that most of the expected fluctuations in the magnetic flux density is actually due to a cross-talk from the temperature and density oscillations associated with MHD waves, implying the opacity effect. However, \cite{2014ApJ...795....9F} simulated a synthetic observation with Fe I 6301.5{\AA} line and suggested that the photospheric magnetic field retrieved from the weak-field approximation provides the intrinsic oscillations in magnetic flux density associated with the wave propagation because of the low magnetic field gradient. This implies the importance of  the vertical magnetic field structure. \cite{1994A&A...291..622C} reported that the difference in vertical gradient of the magnetic flux density observed in the large sunspot's umbrae is about -0.25 G km$^{-1}$. Since this value is close to the condition used in \cite{2014ApJ...795....9F}, there is a strong possibility that observed magnetic fluctuations are intrinsic. \par

The second possible reason is because of the simplified modeling for the seismology. Since the straight cylinder model \citep{2013A&A...551A.137M} does not consider the expanding magnetic shape and the density stratification, there are some differences between the modeling and the observed sunspot. \par

The third possible reason is due to the temperature reduced in the sunspot umbra. The temperature reduced at the umbral photosphere may reduce the amount of $H^-$ ion, which is a dominant absorber in the visible wavelength \citep[e.g.,][]{2002tsai.book.....S}. As a consequence, the line formation layer moves downward and may increase our density estimate to a higher value because of the gravity stratification. Previous studies, such as \cite{2004A&A...422..693M} and \cite{1993A&A...270..494M}, obtained that the magnitude of the Wilson depression is 400-800 km in the umbra, which is sufficiently longer than the scale height at the photosphere ($\sim$ 150 km).\par

At the end, we should note possibilities that a fraction of the derived difference of the energy flux at the two atmospheric layers may not be the dissipated energy. For example, if ascending photospheric waves refract and do not reach the chromosphere, there is an the energy difference, but the energy is not dissipated in the chromosphere. In this study, since slow-mode waves are generally thought to propagate along the magnetic field, the effect of refraction might not be important in sunspot umbrae, where magnetic fields are almost perpendicular to the solar surface. Tracing waves from the photosphere to the chromosphere also helps us understand their true connection. \cite{2015A&A...580A..53L} found photospheric oscillations in sunspot penumbrae that have a slightly delayed counterpart of more defined chromospheric running penumbral waves with larger relative velocities, suggesting that the running penumbral waves propagate upward along inclined magnetic field lines. Inside sunspot umbrae, since waveforms in the photosphere and the chromosphere are not similar to each other because of acoustic cutoff and nonlinear interaction, it is not easy to trace waves like \cite{2015A&A...580A..53L}. For considering acoustic cutoff, Fourier filtering is sometimes applied for investigating the propagating processes \citep{2006ApJ...640.1153C,2009ApJ...692.1211C,2010ApJ...722..131F}. Fourier analyses cannot be applied to nonlinear characteristics (especially seen in the chromosphere), and thus we need to develop such a method in the future for tracing waves from the photosphere to the chromosphere more strictly.

\section{Summary and Conclusions}
  \label{chap:summary}
Using a unique data set from the observations coordinated between {\it Hinode} and {\it IRIS}, we investigated the nature of fluctuations in the temporal evolution of physical parameters observed in the sunspot umbra. After identifying the wave mode of the observed fluctuations, we estimated upward energy fluxes at both the photospheric and transition region layers with the {\it Hinode} and {\it IRIS} satellites. The difference in these energy fluxes is considered as the dissipated energy in the region between the two atmospheric layers. \par
We detected periodic fluctuations in the temporal evolution of the photospheric Fe I, chromospheric Mg II k, and lower transition region Si IV lines. We concluded that there are dominant photospheric standing slow-mode waves with leakages of the high-frequency wave components to the chromosphere. As a quantitative result, we derived $2.0\times10^7$ erg cm$^{-2}$ s$^{-1}$ for the upward energy flux at the photospheric layer and $8.3\times10^4$ erg cm$^{-2}$ s$^{-1}$ for the upward energy flux at the lower transition region. Their difference is larger than the heating rate required at the chromosphere above the sunspot umbra, suggesting that the MHD waves observed at the photosphere can play an important role for heating the chromosphere. However, there is a possibility that the opacity effect can also cause the fluctuations in the temporal evolution of the magnetic flux density. Therefore, what we need to do next is to distinguish $\delta B_{intrinsic}$ and $\delta B_{opacity}$ for better quantitative estimate of the energy flux. \par
{\it Hinode} is a Japanese mission developed and launched by ISAS/JAXA, with NAOJ as domestic partner and NASA and STFC (UK) as international partners. It is operated by these agencies in co-operation with ESA and NSC (Norway). {\it IRIS} is a NASA small explorer mission developed and operated by LMSAL with mission operations executed at NASA Ames Research center and major contributions to downlink communications funded by ESA and the Norwegian Space Centre. We sincerely thank to the {\it Hinode} team and the {\it IRIS} team for providing the coordinated observations used in this article. The authors are supported by MEXT/JSPS KAKENHI Grant Numbers 25220703 (R. K), 25220703, 15H05750, 15H05814 (T. S), 25220703, 26287143, 15H05816 (S. I).

\bibliographystyle{apj}
\bibliography{reference}
\end{document}